\documentclass[iop,apj]{emulateapj}
\pdfoutput=1
\newcommand{\msun}{M_\odot}

\shorttitle{Supernovae in Massive Binaries}
\shortauthors{Hirai et al.}

\usepackage[plainpages=false, colorlinks=true, anchorcolor=blue, linkcolor=blue, citecolor=blue, bookmarks=false]{hyperref}

\begin{document}

\title{The Outcome of Supernovae in Massive Binaries; \\
    Removed Mass, and its Separation Dependence}

\author{Ryosuke Hirai\altaffilmark{1} ,Hidetomo Sawai\altaffilmark{1,2}, and Shoichi Yamada\altaffilmark{1}}
\affil{\altaffilmark{1}Advanced Research Institute for Science and Engineering, Waseda University, 3-4-1, Okubo, Shinjuku, Tokyo 169-8555, Japan}
\affil{\altaffilmark{2}Research Organization for Information Science \& Technology, 1-5-2
Minatojimaminamimachi, Chuo, Kobe, Hyogo 650-0047, Japan}

\begin{abstract}
The majority of massive stars are formed in binary systems. It is hence reasonable to expect that most core-collapse supernovae (CCSNe) take place 
in binaries and the existence of a companion star may leave some imprints in observed features. Having this in mind, we have conducted two-dimensional 
hydrodynamical simulations of the collisions of CCSNe ejecta with the companion star in an almost-equal-mass ($\sim 10{\rm \msun}$) binary to find 
out possible consequences of such events. In particular we pay attention to the amount of mass removed and its dependence on the binary separation. 
In contrast to the previous surmise, we find that the companion mass is stripped not by momentum transfer but by shock heating. Up to $25\%$ of the 
original mass can be removed for the closest separations and the removed mass decreases as $M_{\rm ub} \propto a^{-4.3}$ with the binary separation 
$a$. By performing some experimental computations with artificially-modified densities of incident ejecta, we show that if the velocity of ejecta is fixed, 
the density of incident ejecta is the single important parameter that actually determines the removed mass as  $M_{\rm ub} \propto \rho_{\rm ej} ^{1.4} $. On the 
other hand, another set of simulations with modified velocities of incident ejecta demonstrate that the strength of the forward shock, which heats up the 
stellar material and causes the mass loss of the companion star, is actually the key parameter for the removed mass. 
\end{abstract}


\section{Introduction}

It is known observationally that about half of the observed stars are members of binary or higher order multiple systems \cite[]{rag10,duc13}. The fraction increases 
with the primary star mass and reaches up to 69\% for O stars \cite[]{san12}. These massive stars are supposed to end up exploding as core-collapse 
supernovae (CCSNe). It is hence natural to expect that the majority of CCSNe should take place in binaries. 

Binary interactions may be crucially important for the evolution leading to core-collapse \cite[]{pod04}. In particular, mass transfer will be essential to 
the spectral type of the supernova (SN) explosion. In fact, it is argued that type Ib and Ic SNe mainly occur in interacting binaries \cite[]{smi11}. It has also recently been reported that the late time photometry of iPTF 13bvn, a type Ib SN, cannot be reproduced by single star progenitors, but can naturally be explained by binary progenitor models \cite[]{fre14,ber14}. This may indicate that type Ib SNe are actually occurring in binaries although we had better wait for direct observations of the surviving companions in the future.
The aftermath of the explosion could be no less important for the evolution of the binary system itself as well as of the companion star. It is well known that 
the binary system is disrupted if more than half of the total mass is expelled. This criterion is easily fulfilled when the primary star explodes unless it has lost most of 
its mass before explosion. Even if the amount of mass expelled is less than half the total mass, the system may still be disrupted due to neutron star kicks. 
The companion star will then carry on its life as a single star. 
If the binary survives the SN explosion, it will have a highly eccentric orbit \cite[]{pij12}. Some observational facts of a high mass companion actually surviving the SN explosion are known \cite[]{sew12,gel80}. Most high mass X-ray binaries (HMXBs) and low mass X-ray binaries must have undergone SN at some time in their formation, which means a certain fraction of binaries need to survive from explosion.

Collisions of the supernova ejecta (SNE) with the companion star may also affect the evolution of the latter if the binary separation is small enough. \cite{wlm75} were the
first to estimate the mass removed from the companion star by the impact. They considered momentum conservation in a simple analytical model to express the 
amount of mass removed with a single parameter and estimated that up to 15$\%$ of the companion mass will be ejected.

Numerical computations were also carried out by \cite{frx81}, confirming the results by \cite{wlm75}. Their two-dimensional computational grid of $32\times32$ was 
rather coarse by the current standard. They also assumed a planar shell as a model of SNE, which may not be a good approximation for small binary separations, 
where the spherical geometry of SNE is not negligible. \cite{liv92} followed, assuming a red giant companion of 1 $\msun$. They treated SNE as a 
spherical shell and used a finer mesh, enough to describe the envelope with fine zoning (typically 107 $\times$ 65). They found that almost all the envelope of the red giants were stripped off by SNE. The result may be 
applied not only to type Ib SNe but also to type Ia SNe.

The impact of SNE on companion stars have been better studied for type Ia SNe \cite[]{mar00,pan13,liu13,kas10} in the single degenerate (SD) 
scenario, where a carbon-oxygen white dwarf accretes mass from its low-mass binary companion \cite[]{nom82}. Carrying out numerical calculations with high 
resolutions, these authors placed strong constraints on the structure of companion stars in the SD scenario. 

In this paper, we perform similar simulations, but assuming more massive stars for both the primary and secondary stars, which will be appropriate for CCSNe. 
Unlike for type Ia SNe in the low-mass binary, there is no standard model for CCSNe. In fact, the masses and structures of the primary and secondary stars as 
well as the binary separation are almost free parameters, since there are not many observational constraints. We pay particular attention to the mass removed from 
the massive companion and its dependence on separation. For the analyses of the results, we perform additional simulations of experimental nature, modifying 
the density and/or velocity of SNE artificially. It turns out that these are indeed helpful to pin down which physical quantity is most essential in determining the 
amount of removed mass.

This paper is structured as follows: In Section 2, we describe the models and numerical method we used. The main results are shown in Section 3, and discussions
are given in Section 4. Finally, we summarize our results in Section 5.

\section{Models and Numerical Method}

In this paper we numerically investigate the impacts of the SNE against the companion star in a massive close binary. We employ two-step strategy: 
the explosion of the primary star is computed in spherical symmetry, ignoring the existence of the secondary star, to obtain the structure of the SNE (Step 1); 
the collision of the SNE obtained in Step 1 against the companion star is simulated under the assumption of axisymmetry (Step 2). Details of each step 
are given below.

\subsection{Numerical Method}

We use a two dimensional hydrodynamic code ``yamazakura'' for all computations performed in this paper \cite[]{saw13}. It is a time-explicit Eulerian 
code based on a high resolution central scheme \cite[]{kur00}. Spherical coordinates are employed and axisymmetry but no equatorial symmetry 
is assumed. In Step 1 we further impose spherical symmetry, putting the origin of the coordinates at the centre of the primary star and ignoring its rotation 
as well as the orbital motion and the gravity of the companion star. In Step 2 the existence of the primary star is ignored, with the 
origin of the coordinates coinciding with the centre of the companion star. The symmetry axis is directed to the primary star and rotation of the companion 
star as well as the orbital motion is neglected. The self-gravity of the companion star and SNE in the computational domain is included by solving the 
Poisson equation with the boundary conditions set by multipole expansion. An ideal equation of state with an adiabatic index $\gamma=\frac{5}{3}$ is 
applied throughout. 

\subsection{Binary Parameters}

The masses of the primary and companion stars are set to $M_1=10\msun$ and $ M_2=10.7\msun$, respectively. Note these are not the masses
at birth of the binary system but the values just prior to the SN explosion of the primary star. These values are chosen so that the primary star would be 
massive enough to produce a SN explosion and the mass ratio should be close to unity, in which case the companion star radius $R_2$ could be largest 
for a given separation $a$. 

The $a/R_2$ ratio is expected to be one of the most essential parameters that determine the amount of mass removed by the ejecta \cite[]{wlm75}. 
In order to maximize the impact of SNE on the companion star we choose the separation of binary as small as possible under the restriction that the 
companion should not cause Roche lobe overflow (RLOF). Assuming that the secondary star fills the Roche lobe, we obtain the minimum separation
$a_{\rm min}$ from the Eggleton relation \cite[]{egg83}:
\begin{equation}
 a_{\rm min}=\frac{0.6q^{\frac{2}{3}}+\ln(1+q^{\frac{1}{3}})}{0.49q^{\frac{2}{3}}} R_2,
\label{egg}
\end{equation} 
where $q=M_2/M_1$ is the mass ratio. The period of the Keplerian motion with this separation is approximately five years, which is significantly long compared to 
the characteristic timescale of the collision of SNE with the companion star, $\sim R_2/ v_{\rm ej} \sim \mbox{a few weeks}$. Here $v_{\rm ej}$  is a typical velocity of 
SNE. It is hence justified to neglect the orbital motion in Step 2.

It is one of the main objectives in this paper to find the dependence of the removed mass on the binary separation. For this purpose we have carried out 
several simulations with different separations $a=(1.0,1.1,1.2,... ,2.5) \times a_{\rm min}$.

\subsection{Step 1: Primary Star Explosion} \label{explosionstage}

Since we are not interested in the still unknown SN mechanism, we artificially produce a SN explosion in this study by injecting internal energy into the 
central region of the primary star. Such a procedure is often called the ``thermal bomb'' \cite[]{yng07}. For the progenitor, we use a pre-SN model of Heger 
\& Wellstein 1999 (unpublished), which was computed according to the binary evolution models of \cite{wel99}, in which the ``standard'' Wolf-Rayet mass 
loss rate is reduced by a factor of six and mass transfer is taken into account. The mass and radius of the primary star are $M_1=10\msun$ and 
$R_1=7.9\times10^{5}{\rm km}$, respectively, just prior to the SN explosion. It has totally lost its H layer, and is a helium star since it has undergone a significant amount of mass loss. 

We cover the primary star with a one-dimensional spherical grid. Sixteen-hundred grid points are non-uniformly distributed  in the radial direction. The inner boundary is 
chosen so that the central $1.4{\rm M_\odot}$, corresponding to a neutron star, should be excluded from the computational region. The outer 
boundary is set at approximately 40 times the stellar radius. $10^{51}$ erg of internal energy is added to the inner 10 grids. Self-gravity is included in this 
computation, although its effects are limited. We impose the reflective condition at the inner boundary whereas the outgoing boundary condition was used at the outer boundary.

After the shock wave breaks out of the stellar surface, the SNE continues to expand, converting internal energy to kinetic energy. Since the hydrodynamic code 
cannot handle vacuum, we artificially fill the volume outside the star with matter, which is so dilute that it should not affect the dynamics. When the SNE reaches the
outer boundary of the computational domain, it expands almost self-similarly. We record the density, pressure, velocity at an arbitrarily chosen point near the outer 
boundary as a function of time so that we could use them in Step 2 as described shortly.

\subsection{Step 2: Collisions of SNE on the Companion Star} \label{ejectastage}
 
\begin{figure}
    \plotone{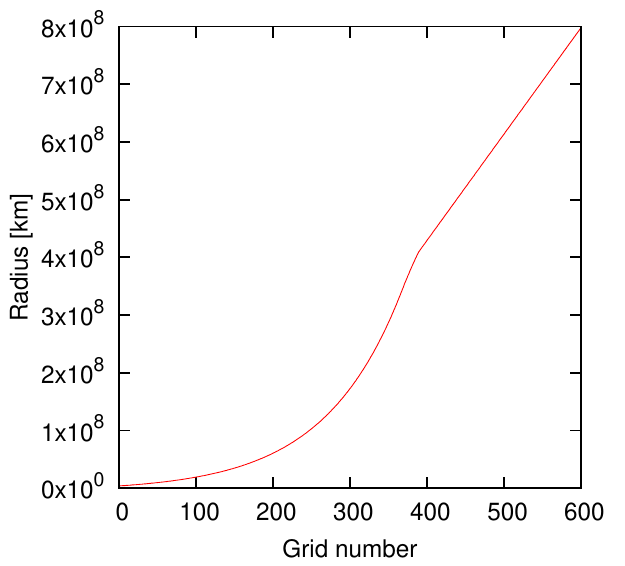}
    \caption{Locations of the radial grid points used in Step 2.\label{gridstructure}}
  \end{figure}

As mentioned above, we place the companion star at the centre of a two-dimensional spherical mesh, whose axis is aligned with the line joining the centre of 
the two binary components. Neglecting rotation of the companion star and the orbital motion, we assume axisymmetry in this step. The computational domain 
extends from the radius of $r=4\times10^{6}{\rm km}$ to $r = 8\times10^{8}{\rm km}$, which is roughly twice the radius of the companion star. The central portion that contains $2.79\msun$ is cut out to avoid too
severe CFL conditions. We treat it as a point gravitational source and we impose a reflective inner boundary condition. The computational region is divided into $600\;(r) \times 180\;(\theta)$ 
zones non-uniformly in the radial direction and uniformly in the $\theta$ direction. It is important to put the radial grid points appropriately in order to keep the star in 
hydrostatic equilibrium. We deploy at least 20 mesh points in each density scale height inside the star. On the other hand, we distribute the radial grid points
uniformly outside the star, where we again put a dilute gas for numerical convenience. The locations of the grid points determined this way are shown in Figure \ref{gridstructure}. 

For the companion star we adopt the non-rotating s11.0 model from \cite{heg02}, since no information is available on the companion in Heger \& Wellstein 1999 (unpublished). 
This companion star model gives a mass ratio close to unity and gives the smallest value to $a_{\rm min}/R_2$ among available stellar models. In fact, its radius of the companion star is $R_2 =4.08\times10^{8}{\rm km}$ and the $a/R_2$ ratio used in this study ranges from 2.7 to 6.7. Since it is a pre-SN model, i.e., the star is near the end of its evolution and has a loosely bound envelope, so the results of our simulation possibly serve as an upper limit for the amount of mass stripped in CCSNe of primary stars. It is true that the companion model is actually a single star model and not consistent with the primary star model, and may not be a realistic binary system. But massive binaries that will produce a CCSN are very much  diverse and our understanding of their evolutions is not yet complete. Moreover, binary-progenitor models available to us for the moment are not many, either. In this paper, we hence  focus on the systematics of interactions between SN ejecta and the companion star, paying particular attention to the binary separation. Our simulations are experimental in this sense. Exploration of the dependences on the mass, structure and composition of the companion star for more realistic models will be left for future works.

\begin{figure}
 \plotone{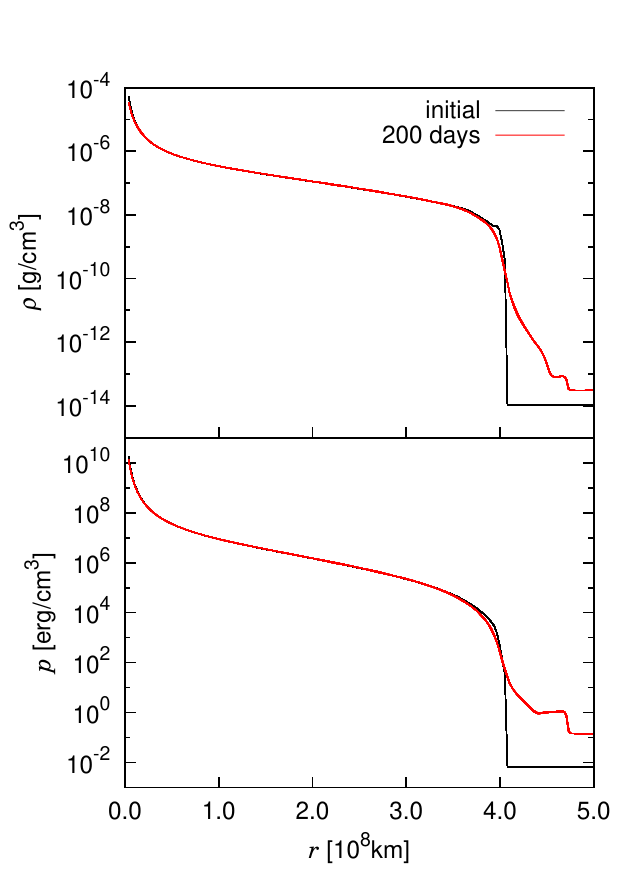}
 \caption{Density ({\it upper}) and pressure ({\it lower}) profiles of the companion star model. The initial profile (black line) is compared with the profiles of the star after being relaxed on the two-dimensional computational grid for 200 days (red line).\label{equil}}
\end{figure}

As mentioned above, the companion star is adopted from pre-SN models and its central core has already started to collapse. The computational domain excludes the central collapsing core, however. It is confirmed that the remaining envelope is almost in hydrostatic equilibrium and does not change significantly for at least 200 days after mapped onto the two-dimensional computational grid, which is shown in Figure \ref{equil}.

SNE is injected from one hemisphere of the outer boundary with its spherical geometry being taken into account properly. We estimate the density, pressure and velocity of SNE from 
the data obtained in Step 1, assuming that the SNE continues to expand self-similarly from the outer boundary of Step 1 to that of Step 2. The free boundary condition is 
applied to the other hemisphere of the outer boundary. Simulations start from the time, when the outermost part of the SNE touches the surface of the companion star.

To follow Lagrangian motions of fluid elements in our Eulerian hydrodynamics code and find which portions of the companion star are stripped by SNE, we deploy 
tracer particles in the companion star as well as in the SNE. The tracer particles are moved with the fluid velocities at their instantaneous locations, which are obtained by the
hydrodynamical computations, and do not affect the dynamics of SNE and stellar matter. Each particle is labelled whether it originates from the star or ejecta and whether it 
escapes out of the computation domain before the termination of simulations or not. These information are extremely useful to understand how the mass-stripping takes place in 
the collision of SNE and the companion star. 

\section{Results}
 
 In the following the numerical results for each step are given in turn.

\subsection{Step 1: Primary Star Explosion}

We artificially initiate the SN explosion in the primary star by depositing thermal energy near the inner boundary. The pressure excess then drives the shock wave outwards
into the envelope. The deposited energy of $10^{51}{\rm erg}$, the canonical value for CCSNe, is large enough to expel the outer envelope, since its binding energy is much 
smaller. The shock wave continues to propagate outward unhindered and breaks out of the stellar surface after 30 s. 

\begin{figure}
  \plotone{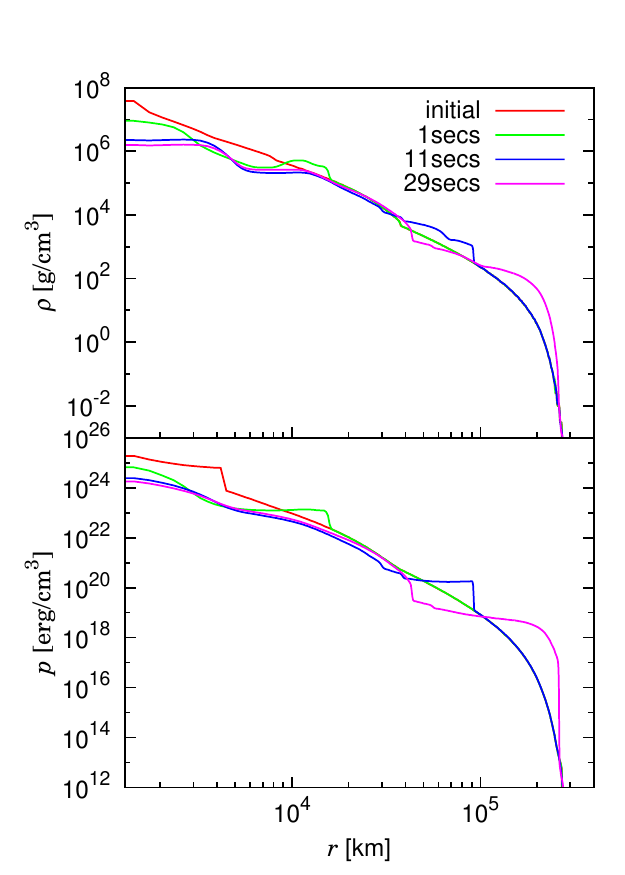}
  \caption{Shock propagation through the primary star. Density and pressure distributions are shown on the top and bottom panels, respectively. The CO/He boundary lies at 
  $r=3.8\times10^4$km, where there is a ``hump'' in the initial distribution.\label{shock}}
 \end{figure}

Figure~\ref{shock} shows the propagation of the shock 
wave in the stellar envelope. It is clear that a dense shell is formed after the shock wave. It is also discernible that a reverse shock is produced when the forward shock passes
through the boundary of C/O and He. It is known that the deceleration of matter by negative pressure gradients then induces a Rayleigh-Taylor instability in the vicinity of the CO/He
boundary. We ignore such multi-dimensional effects in the present study, anticipating that the dynamics of collision of the SNE and the companion star, particularly the removed
mass, will not be affected severely by the small-scale perturbations.

\begin{figure*}
  \epsscale{.8}
  \plotone{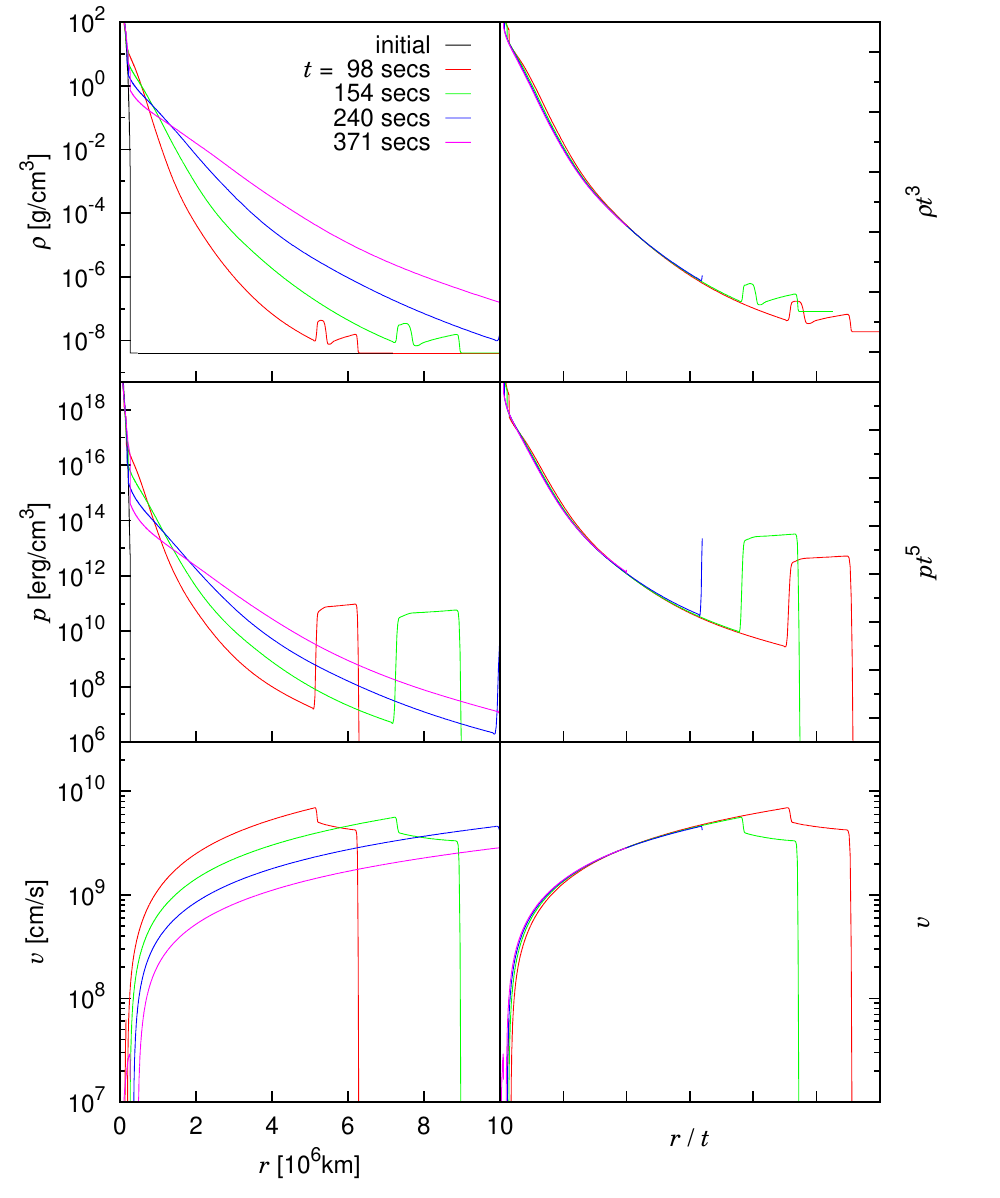}
  \caption{Spatial distributions of the SNE at various times. Left panels show the actual spatial distributions, whereas right panels display those rescaled by the time since 
  SN explosion.\label{profile}}
 \end{figure*}

After the shock break-out, the stellar envelope starts to expand into interstellar matter (ISM) as an SNE. It is initially accelerated by its own pressure, converting thermal energy to kinetic 
energy. When most of the thermal energy is consumed, the SNE expands ballistically until it sweeps up an amount of ISM comparable to the SNE itself and the Sedov-Taylor 
phase commences. This is indeed observed in our simulation as shown in Figure~\ref{profile}. In the left panels the profiles of density, pressure and velocity are shown for different 
times after the shock break-out whereas in the right panels the rescaled distributions are compared. Self-similarity of the expansion is evident from the latter comparison. We assume 
in this study that this free expansion of SNE continues until it reaches the companion star.

 \begin{figure*}
  \epsscale{.8}
 \plotone{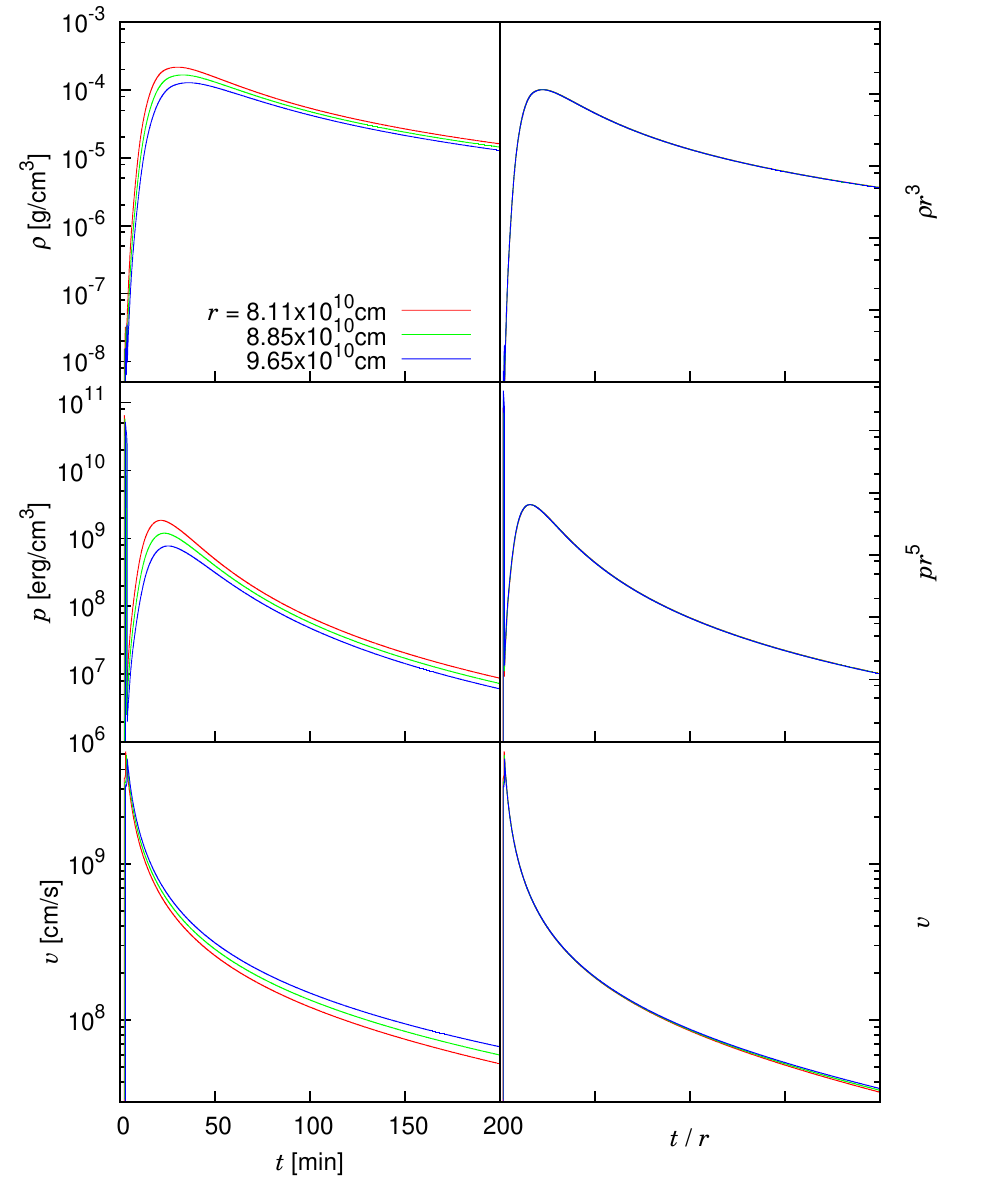}
 \caption{Profiles of density, pressure and velocity of the SNE. Left panels show the actual profiles at distances $r = 8.11, 8.85, 9.65\times10^{10}$cm from the exploding star. 
 Right panels display the same quantities but scaled by the distance. We can see that the three lines agree almost exactly, which indicates a self-similar expansion.\label{explosion}}
 \end{figure*}

The computation of Step 1 is terminated when most of the SNE leaves the computational domain. The temporal evolutions of density, pressure and velocity are recorded at distances 
$r = 8.11,\: 8.85,\:  9.65\times10^{5}$km from the centre of the exploding star. They are shown on the left panels of Figure \ref{explosion}. As the dense shell of SNE passes through 
these points, the density and pressure rise rapidly and decline gradually. The self-similarity in the expansion of SNE is also reflected in these profiles as demonstrated on the right panels, 
where we rescale the profiles by the distance $r$. We can see that the three lines almost exactly agree after this rescaling, which indicates that the SNE expands in a self-similar fashion. 
We utilize these rescaled profiles to set the initial and boundary conditions in Step 2.

\subsection{Step 2: Collisions of SNE on the Companion Star}

\subsubsection{Dynamics of Collisions} \label{dynamics}

\begin{figure*}
   \epsscale{1.20}
  \plotone{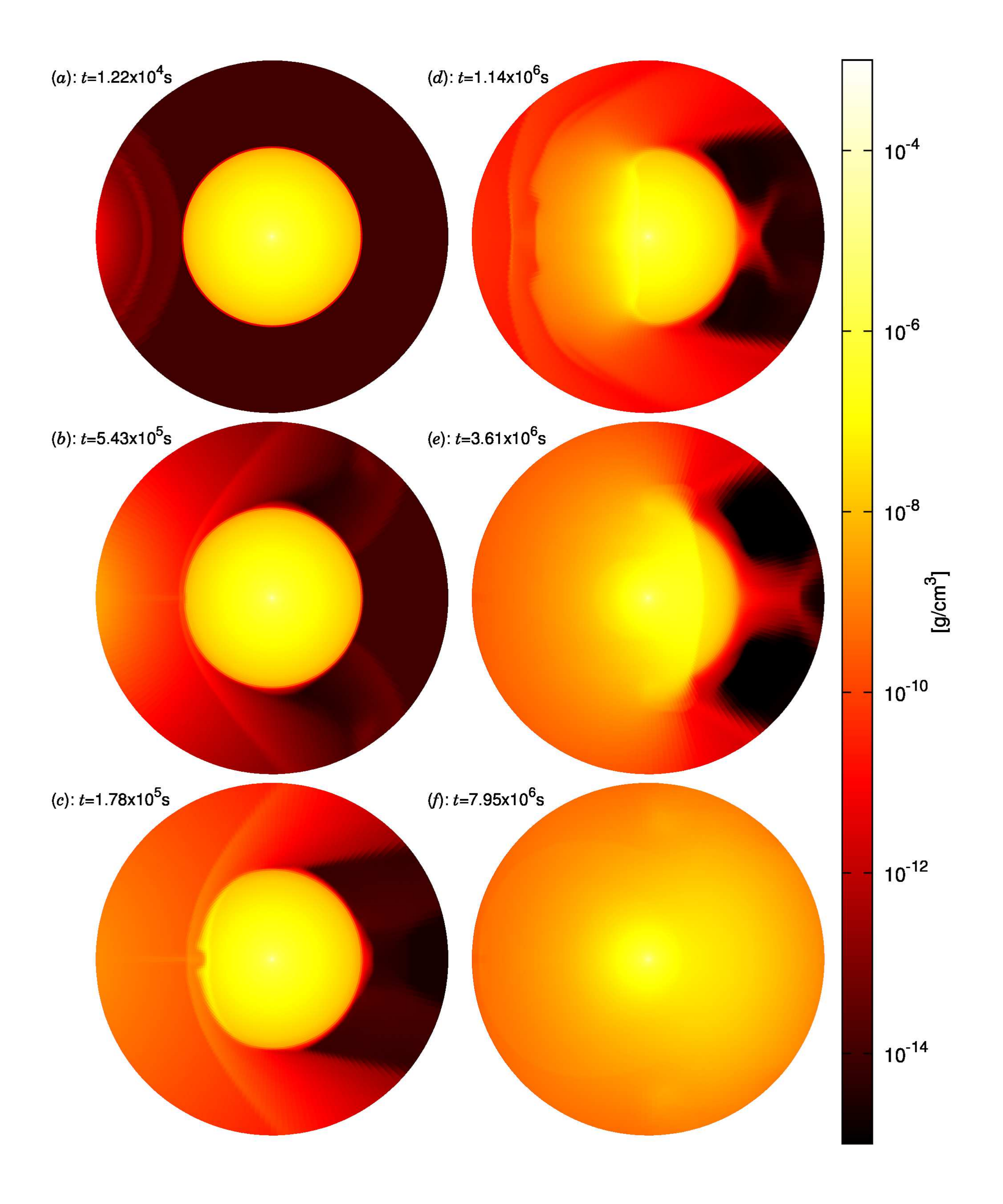}
  \caption{Density profiles at Step 2. The time elapsed since SN explosion is indicated for each panel. The radius of each circle is $8\times10^{13}$cm. The colour bar is in log scale.\label{density}}
  \end{figure*}

\begin{figure*}
  \plotone{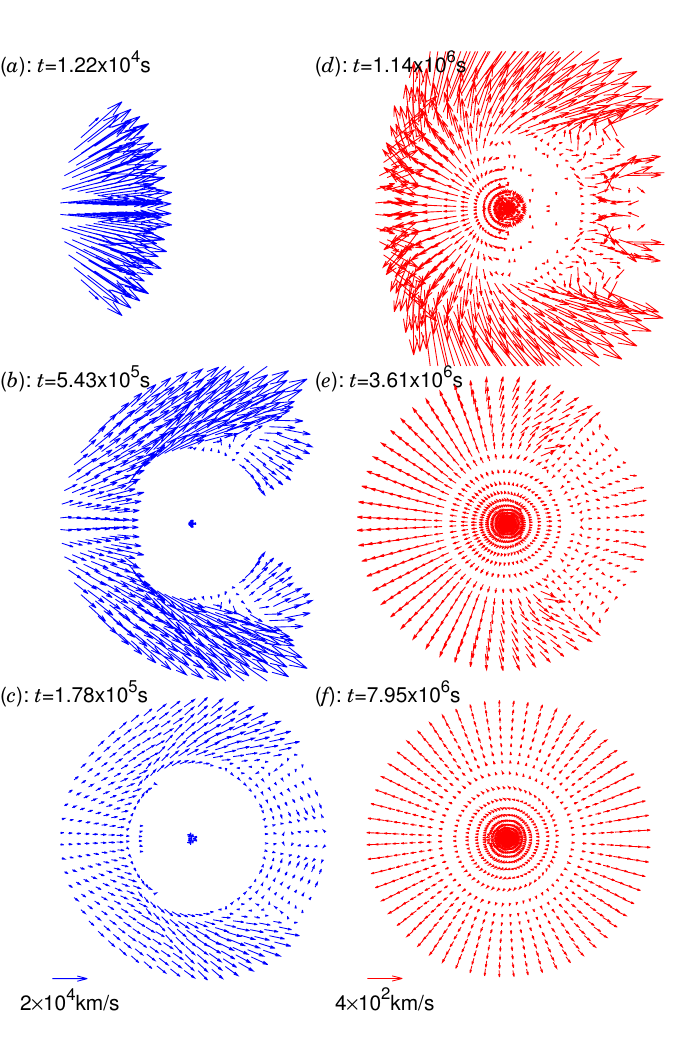}
  \caption{Velocity vectors at Step 2. Note that red arrows on the right panels have lengths amplified by a factor of $50$ compared with blue arrows on the left panels.  \label{pltv}}
  \end{figure*}

We first describe the dynamics of the collisions of the SNE and the companion star for the base model, in which the binary separation is set to $a=a_{\rm min}$. The results for other models with 
different binary separations will be presented in Section~\ref{secsepdep} but qualitative features are common. The density and velocity profiles are shown at different times $t$ elapsed 
since the SN explosion of the primary star in Figures ~\ref{density} and \ref{pltv}, respectively. The top left panels in these figures correspond to the initial state for Step 2. Note that the lengths of vectors 
are scaled for visibility and amplified by a factor of $50$ for the right panels compared to the left panels because the velocity becomes quite small at these late stages.

We can see that as the SNE hits the companion star surface, a forward shock is formed and penetrates through the star. This shock wave heats the stellar matter as it propagates, and eventually 
reaches the other side of the star. The shock decelerates as it climbs the steep density gradient of the star toward the centre of the companion star. It then accelerates at the latter half, 
climbing down the density gradient from the centre to the stellar surface and depositing much smaller energies. Some of the heated matter starts to expand (panels (d) and (e)). 
A reverse shock is also formed and becomes a bow shock in the SNE. It is almost stationary at the beginning as the ejecta flows past (panels (b) and (c)). Once the ejecta density starts 
to decrease, however, it begins to expand leftwards into the SNE and eventually gets out of the computational domain (panels (d) and (e)). After the forward shock penetrates the entire star, 
the whole star slowly expands almost spherically to reach a new equilibrium (panel (f)). 

The SNE does not strip the stellar matter directly as assumed in the previous papers \cite[]{wlm75}. This is mainly due to the presence of the bow shock, which deflects the ejecta away. 
Most of the removed mass is instead ablated off with the internal energy deposited by the forward shock and is also pulled off by the rarefaction wave generated by the reverse shock. These 
results are consistent with the previous findings for low mass companion stars in type Ia SNe \cite[]{mar00,pan13,liu13,kas10}. The shock break-out of the opposite side of the companion star 
also produces some mass ejections there.

\begin{figure*}
  \epsscale{.82}  
  \plotone{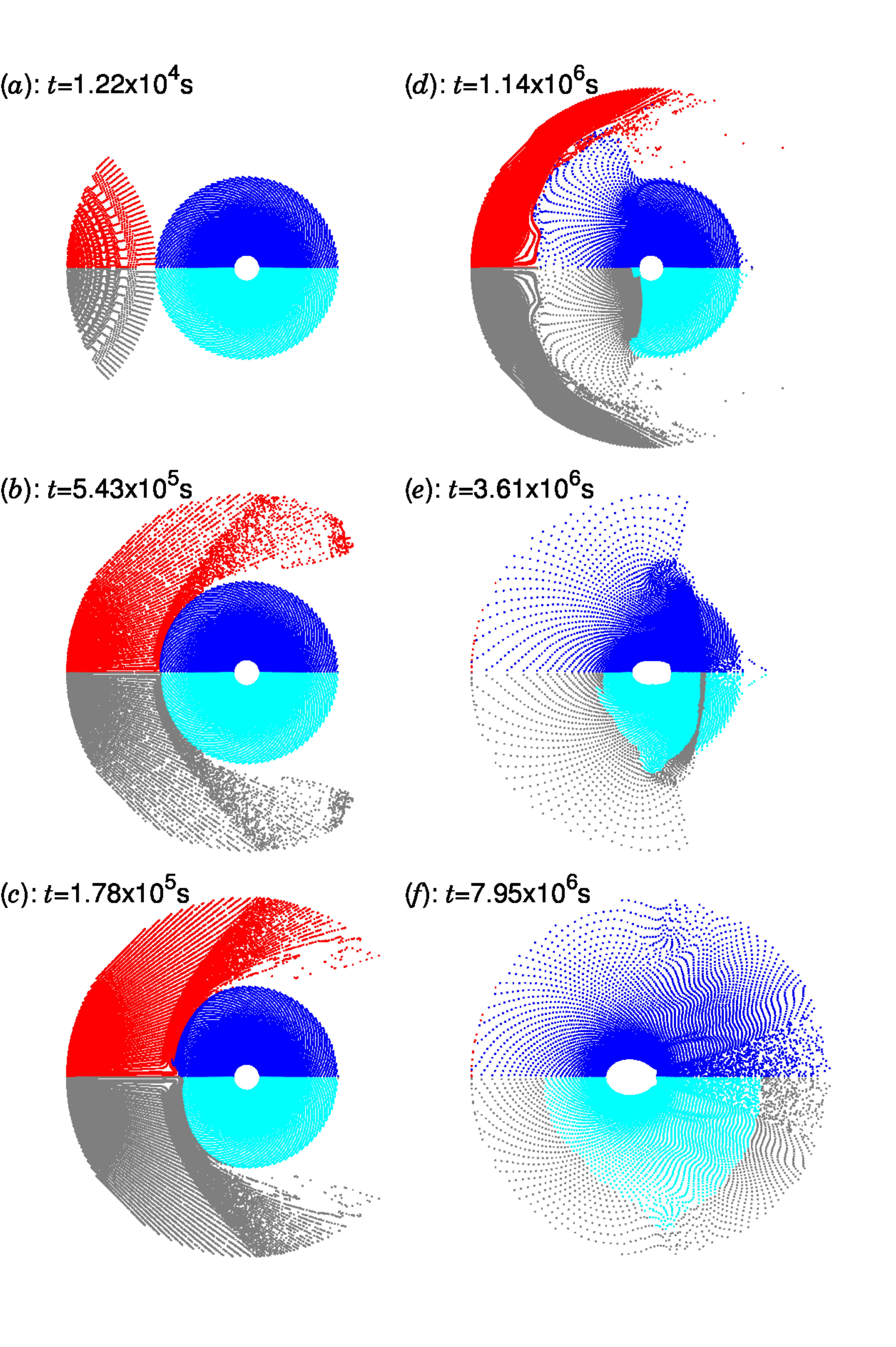}
  \caption{Tracer particles at various times. The upper half of each panel indicates in colour whether the particles originate from the SNE (red dots) or the companion star (light blue dots). 
  The lower half shows in colour whether the particles are bound to the star (blue dots) or not (grey dots).\label{ptc}}
  \end{figure*}

The above statements are more clearly vindicated in Figure \ref{ptc}, where the positions of tracer particles are displayed at the same instants as in Figures \ref{density} and \ref{pltv}. In the upper 
half of each panel the particles originating from the star (blue) are distinguished from those from the SNE (red) by colours. The lower half, on the other hand, shows whether it is "bound" to the 
companion (light blue) or not (grey). The definition of the boundedness will be given in the next subsection in detail. We can see that as the shock propagates through the star, the heated matter 
becomes unbound initially. Although the central region quickly becomes bound again thereafter, the outer part remains unbound and expands almost radially, an indication of thermal ablation. 
On the opposite side of the star, only the low-density matter near the stellar surface becomes unbound by the shock passage because the shock is weakened by that time due to the pull-back 
by the rarefaction wave. The central bound region then restores a spherical shape after a long time. The unbound region continues to expand, on the other hand, and eventually reaches the 
edge of the computational domain. Since we cannot determine for sure the fate of the particles that have left the computational region, we stop the simulation there. This constraint led to the termination time of 122 days for the closest separation model and that down to 105 days for the models with the farthest separations.

Figure \ref{ptc} also points out that there is almost no ejecta contamination going on on the companion star surface. At the end of the simulation, all bound particles are of companion star origin. This indicates that we will not be able to observe imprints of the primary star on the surviving companion after SN.

\subsubsection{Removed Mass} \label{removedmass}

\begin{figure}
  \plotone{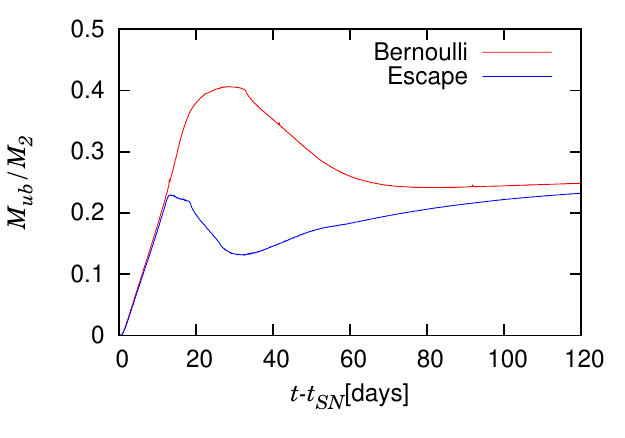}
  \caption{Removed mass as a function of time for the base model with $a=a_{\rm min}$. It is evaluated by two different criteria: the escape criterion (blue line) and the Bernoulli criterion (red line). 
  See the text for details. \label{comparecriterion}}
 \end{figure}

 In previous studies on type Ia SNe \cite[]{mar00,pan13,liu13}, the stripped mass is estimated from the total energy defined locally as 
 \begin{eqnarray}
  \epsilon_{\rm tot}=\frac{1}{2}v^2+\epsilon_{\rm int}+\phi,
 \end{eqnarray}
where $v$ is velocity, $\epsilon_{\rm int}$ the specific internal energy and $\phi$ the gravitational potential. If this energy is positive, which is equivalent to having a velocity exceeding the local escape 
velocity, the fluid element there is judged to be unbound. We shall call this the ``escape criterion''. The removed mass is the sum of these unbound masses. It should be noted, though, that 
the removed mass defined this way is time-dependent particularly at early times because the pressure inside the star is still large and interactions among fluid elements cannot be neglected. 
It takes rather long for the removed mass calculated by this criterion to settle to the final value as shown in Figure \ref{comparecriterion}, in which the temporal evolution of the removed mass is
shown. 

We find that an alternate criterion is more useful to estimate the removed mass. It is based on Bernoulli's principle, which claims under certain conditions that particles along a streamline 
conserve the following quantity
 \begin{eqnarray}
  \epsilon_{\rm ber}=\frac{1}{2}v^2+\epsilon_{int}+\frac{p}{\rho}+\phi,
 \end{eqnarray}
where $p$ is the pressure density and $\rho$ the mass density. Strictly speaking, the principle is valid only in steady systems. Regardless, we assume that a fluid element is unbound if 
$\epsilon_{\rm ber}$ is positive. Note that $\epsilon_{\rm ber}$ should eventually approach $\epsilon_{\rm tot}$ when the pressure becomes negligible. The removed mass is defined as the sum of masses of 
unbound fluid elements determined this way. We refer to this scheme as the ``Bernoulli criterion'' in the rest of this paper. 

In Figure \ref{comparecriterion} we compare the removed masses obtained by the two criteria as a function of time for the base model with $a=a_{\rm min}$: the escape criterion is shown as the 
blue line whereas the Bernoulli criterion is displayed with the red line. It is evident that both lines approach the same asymptotic value, indicating that the two criteria give identical results 
when computed for long enough. More importantly, the red line reaches the final value earlier than the blue line, that is, the Bernoulli criterion is more efficient to estimate the removed mass. This is 
indeed important, since the computing resource is limited and in some models we cannot afford to run simulations long enough to see the settlement of $\epsilon_{\rm tot}$.

 The shock heating of the matter facing the primary star, the main heating phase, lasts approximately 20 days. It takes about 70 days for the removed mass by the Bernoulli criterion to settle to 
 the asymptotic value, which is the time scale for the forward shock to reach the other side of the star. This is merely a few \% of the orbital period and supports our neglection of the orbital motion 
 in this study. At this time, the whole star starts expanding almost spherically and the expansion carries on for a few more months, where the effects of orbital motion and possible stellar rotation may 
 become non-negligible.

The amount of mass removed reached up to 25$\%$ of the original companion mass in the base model. This is a significant fraction for the later evolutions of both the companion star and the 
binary system itself. It can be regarded as an alternate scenario of mass loss from high mass stars other than binary interaction or stellar winds. In the base model we assume the smallest possible 
binary separation and the removed mass will be largest. The next question is obviously how the removed mass depends on the binary separation.
 
 \subsubsection{Numerical Convergence} \label{secnumconv}

 Before proceeding to other models with different binary separations, we check here whether the spatial resolution employed in the simulation is appropriate. For this purpose 
 we ran models with the same physical setup but with different spatial resolutions. The number of grid points was changed separately for $r$ and $\theta$ directions. They are 
 summarized in Table \ref{convtesttable}. 
 
\begin{figure}
 \epsscale{1.2}
 \plotone{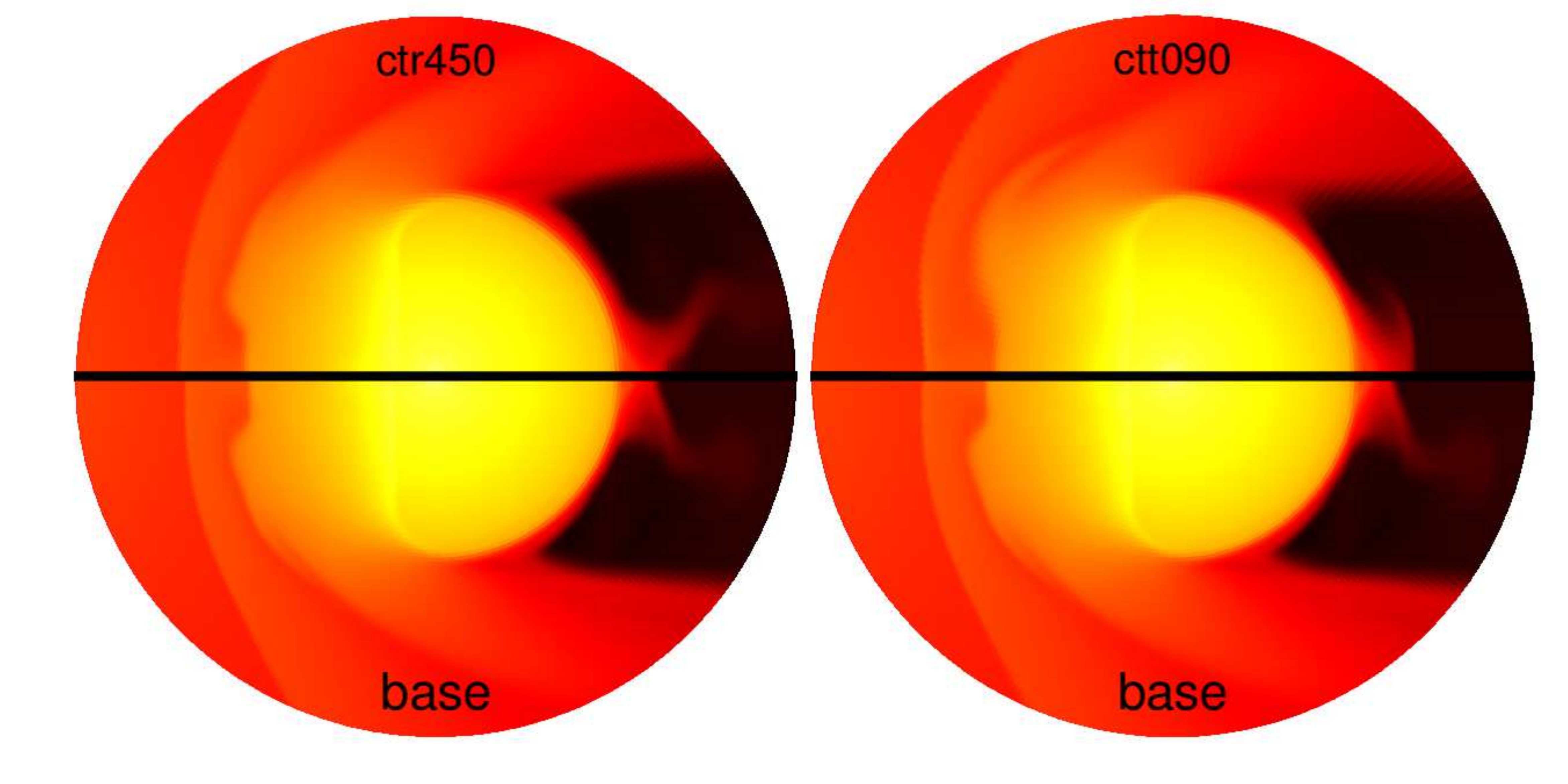}
 \caption{Density distributions for different resolutions. Snapshots at the time $t=8.5\times10^5$ from models ctr450 ({\it left}) and ctt090 ({\it right}) are compared with 
 that from the base model. The grids used for each model are given in Table \ref{convtesttable}. \label{convcheck}}
\end{figure}

 The dynamics of these test runs are compared with the base model in Figure \ref{convcheck}. The flow patterns hardly show any difference for the models with different grids in the $r$ direction. 
 On the other hand, as the grid becomes coarser in the $\theta$ direction, the shock waves are slightly broadened, while complex flow patterns particularly behind the companion star are changed
 even qualitatively. The removed mass, which is our main concern in this paper, is hardly affected, however. Its values are listed on the table for each model. Note that the number of tracer particles 
 is unchanged in these experimental computations. Although there is a clear trend that the removed mass decreases as the number of grid points increases, the variation is small, $\sim 1\%$, and 
 does not affect our conclusion. We hence believe that the resolution employed in this study is appropriate. 

 \begin{table}
  \centering
  \caption{Spatial Resolutions Adopted in Convergence Tests}
   \begin{tabular}{lccc}
    \hline\hline
    Model  & $r$-grids\tablenotemark{a}  & $\theta$-grids\tablenotemark{a} & $M_{\rm ub} (\msun)$\tablenotemark{b} \\ \hline
     Base  & 600        & 180            & 2.594 \\
    ctr550 & 550        & 180            & 2.620 \\
    ctr500 & 500        & 180            & 2.627 \\
    ctr450 & 450        & 180            & 2.639 \\
    ctt150 & 600        & 150            & 2.629 \\
    ctt120 & 600        & 120            & 2.665 \\
    ctt090 & 600        & \ 90           & 2.704 \\
    \hline

   \end{tabular}\label{convtesttable}

   \footnotetext{$r$-grids and $\theta$-grids show the number of grids used in each direction.}
   \footnotetext{$M_{\rm ub}$ is the removed mass.}

 \end{table}

 \subsubsection{Separation Dependence}\label{secsepdep}

In the base model, the binary separation is assumed to be $a = a_{\rm min}$. This is the smallest possible value and is most likely to be too small. We have hence performed the same simulation 
for several models varying the separation as $a=(1.1, 1.2, ... , 2.5)\times a_{\rm min}$ to see the dependence of removed mass on separation. As the binary becomes wider, the density of SNE
at the impact against the companion star is decreased; the SNE is more extended specially and the collision lasts longer; since the solid angle of the companion star is also decreased, the total
energy contained in the portion of SNE that collides with the companion star is reduced proportionately; the curvature of the dense shell in the SNE becomes less pronounced. All of these effects 
but the last one will tend to decrease the amount of the removed mass. The smaller curvature of the SNE may lead to smaller deflections by the bow shock and contribute to some direct stripping.
 
\begin{figure*}
 \epsscale{1.1}
 \plotone{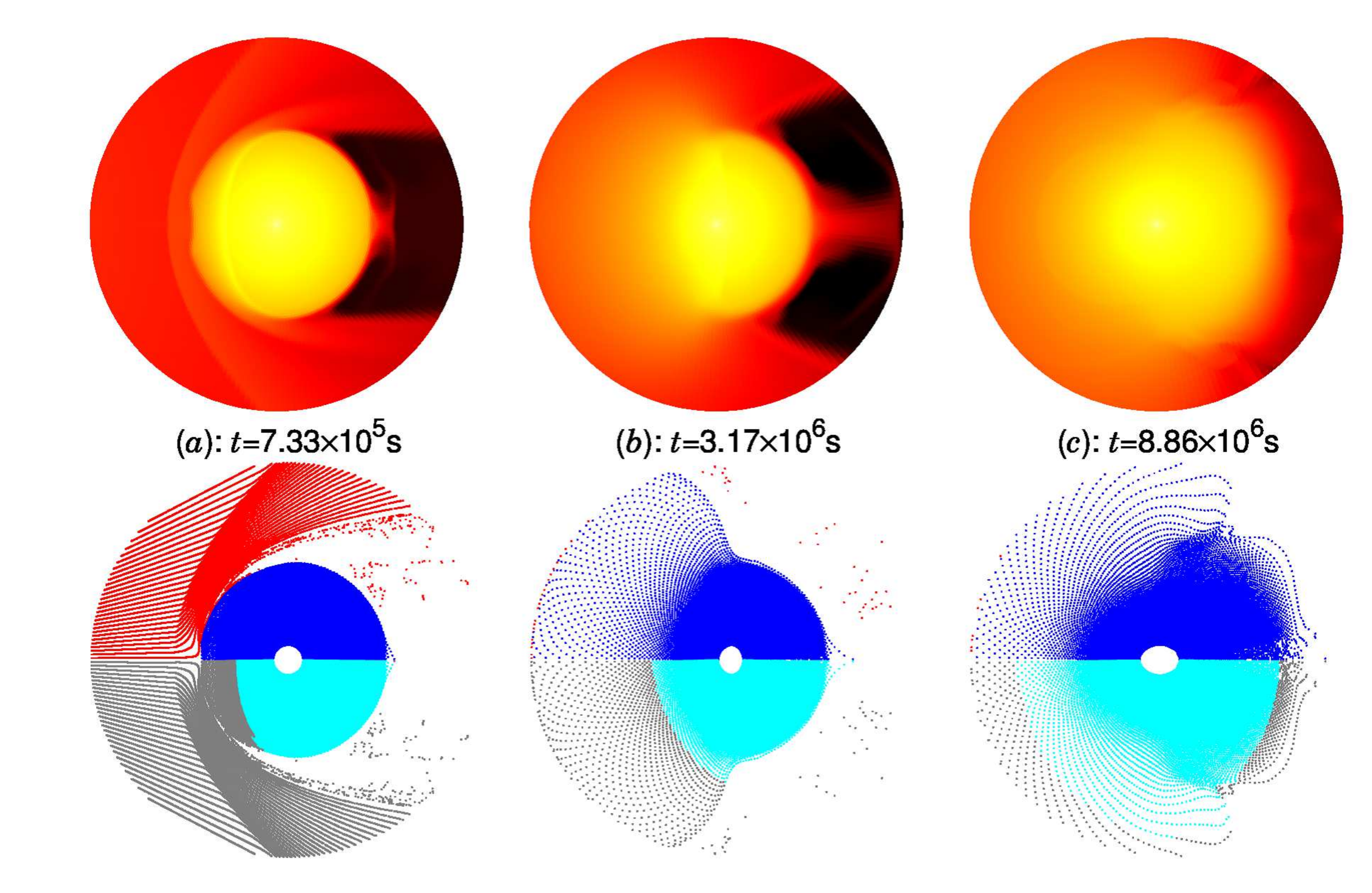}
 \caption{Time evolutions of the density distribution ({\it upper panels}) and tracer particle distribution ({\it lower panels}) for the model with $a=1.5\times a_{\rm min}$ . The same colour bar is used for 
 density as in Figure \ref{density}. The particles are classified with the same colours as in Figure \ref{ptc}. \label{sep15}}
\end{figure*}

\begin{figure*}
 \epsscale{1.1}
 \plotone{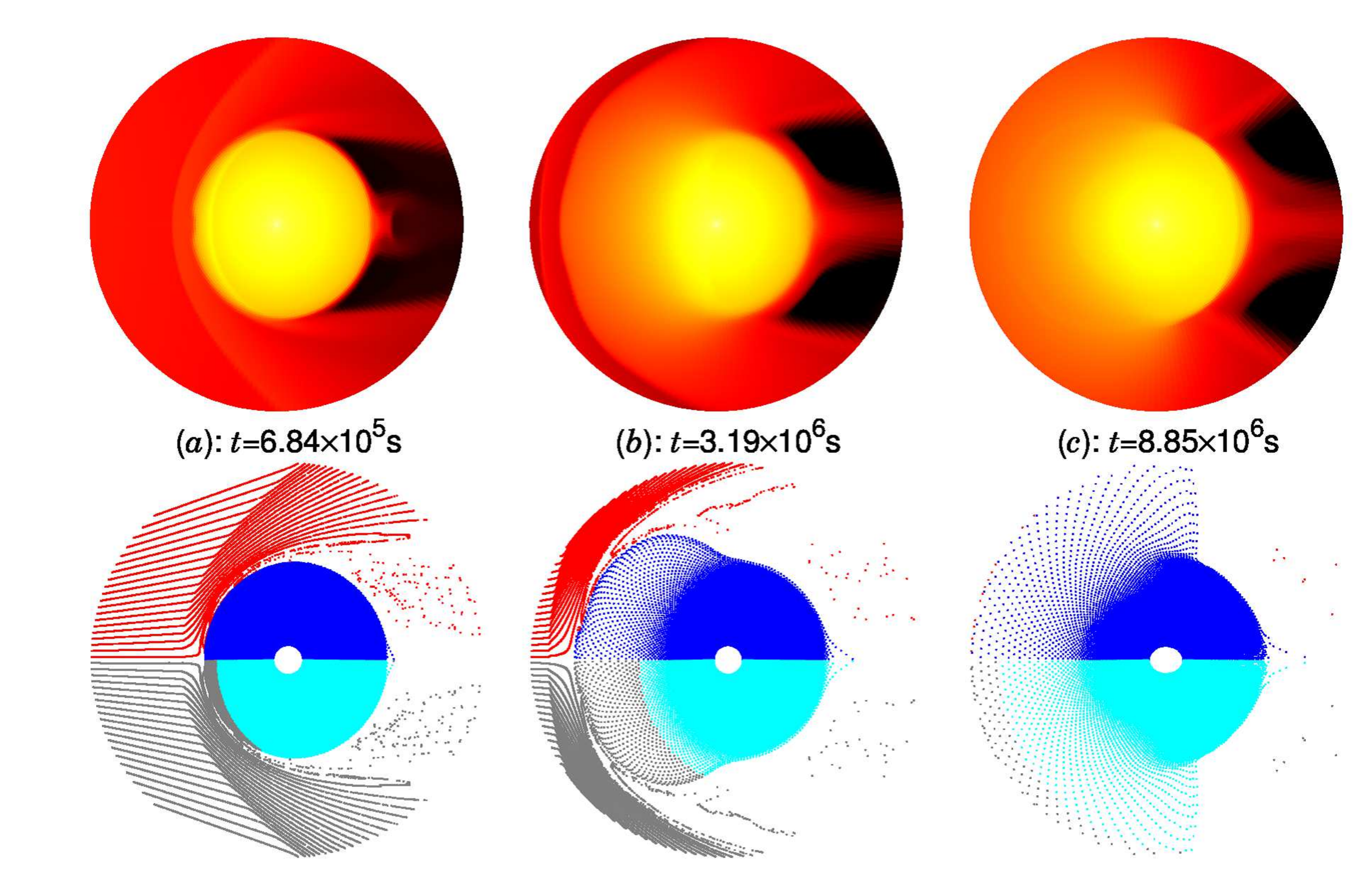}
 \caption{Time evolutions of the density distribution ({\it upper panels}) and tracer particle distribution ({\it lower panels}) for the model with $a=2.0\times a_{\rm min}$. The same colour bar is used for 
 density as in Figure \ref{density}. The particles are classified with the same colours as in Figure \ref{ptc}. \label{sep2}}
\end{figure*}

 For other models with wider separations, forward and reverse shocks are formed when the SNE hits the surface of the companion (panel (a) of Figures \ref{sep15} and \ref{sep2}) as was 
the case for the base model. The forward shock has smaller pressure and velocity, since the incident ejecta energy density is smaller. Since it takes the forward shock longer to sweep 
the entire star, the hemisphere of the companion star that faces the primary star and is shock-heated earlier has a highly extended envelope compared to the opposite hemisphere that 
is heated later, making the star take a very asymmetric shape (panel (b) of Figures \ref{sep15} and \ref{sep2}). Since the shock heating is significantly weaker in the wide binaries, only the 
initially shocked regions become unbound and the total removed mass is smaller. As the forward shock reaches the opposite side of the companion star (panel (c) of Figures \ref{sep15} and \ref{sep2}), much smaller amounts of matter become unbound near the surface.

 \begin{figure}
  \plotone{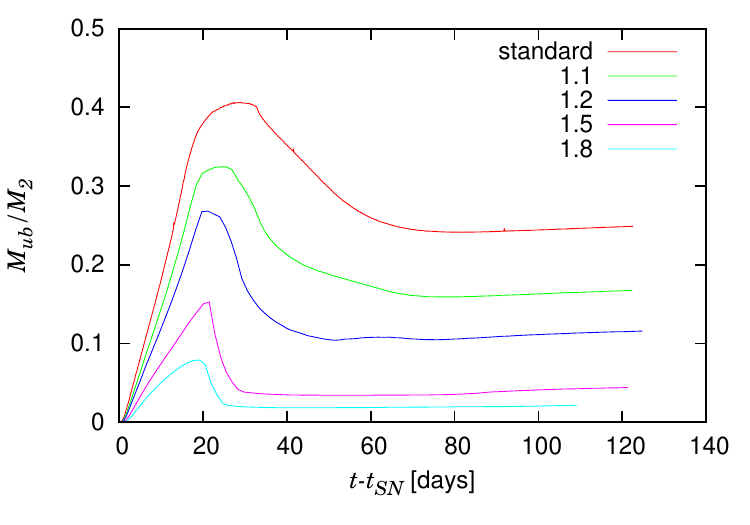}
  \caption{Time evolutions of removed mass are shown for several models. The removed masses are evaluated by the Bernoulli criterion. The time is measured from SN explosion. The termination times for the simulations shown here are 122, 122, 125, 121, 109 days respectively.\label{timemub}}
 \end{figure}

\begin{figure}
 \plotone{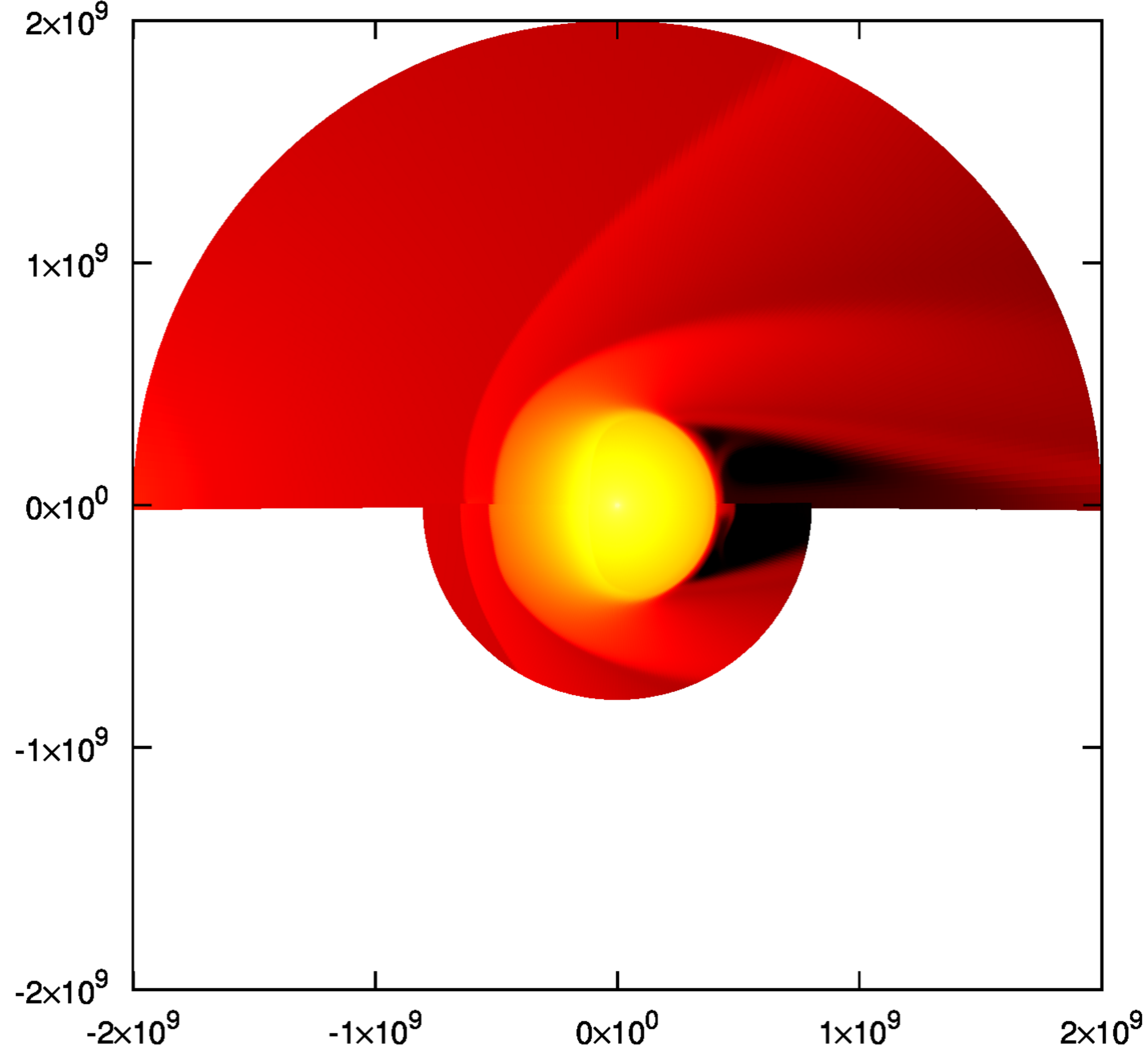}
 \caption{Comparison of density distributions in the larger-domain run ({\it upper half}) and the original domain run ({\it lower half}) for the $a=2.0\times a_{\rm min}$ model. The axes are in units of kilometres. The time is $2.22\times10^6$ s.\label{wide}}
\end{figure}

\begin{figure}
 \plotone{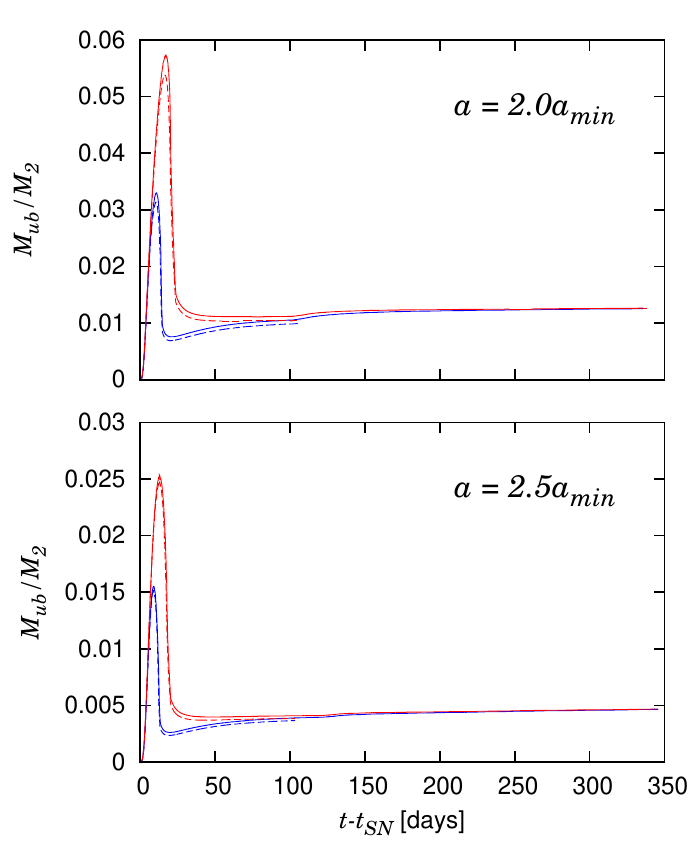}
 \caption{Time evolutions of the removed mass for the larger-domain runs. Results for the $a=2.0\times a_{\rm min}$ model is shown on the upper panel, and $a=2.5\times a_{\rm min}$ on the lower panel. Red lines show the removed masses estimated by the Bernoulli criterion, and blue lines by the escape criterion. Results of the original domain are also plotted with dashed lines.\label{mublb}}
\end{figure}

The removed masses are calculated in the same way as in Section \ref{removedmass} and are shown as a function of time in Figure \ref{timemub} for some representative models. Other 
models have similar features. Although the removed masses tend to asymptotic values within a few months in all models, those with wider separations settle at earlier times. It will also 
take more time until the companion regains mechanical and thermal equilibrium in these models. 

For the models with the widest separations, the heated but still bound matter reaches the outer boundary before the forward shock reaches the other side of the companion star. This may lead to errors in the estimation of the unbound mass, since the heating has not finished, and our Bernoulli criterion may not be applicable yet. To see the possible effects of the size of the computational domain, we performed test calculations with wider computational regions. We set the outer boundary to be $2\times10^9$ km in these simulations, which is $2.5$ times the original size. The inner boundary and spatial resolution inside the companion star were chosen to be the same, while we increased mesh size in a geometric way outside. With this wider domain, we can continue the simulations longer. 

Figure \ref{wide} shows a snapshot of the simulation with the new region, which is compared with the result for the original region. Although the position of the reverse shock is slightly different, there are no qualitative differences in the overall behaviour. The expansion of the bound region continued throughout our simulation, and even exceeded our wider boundary. This is approximately five times the original stellar radius and will keep on expanding for a while. The time evolution of the unbound mass is shown in Figure \ref{mublb}. Despite the fact that the expansion has not ceased, the unbound mass settles to a constant value at fairly early stages. Results for the original and larger domains differ by $\sim16\%$ for the $a=2.0\times a_{\rm min}$ case, and $\sim18\%$ for the $a=2.5\times a_{\rm min}$ case. This indicates that the size of the computational domain does not affect the removed mass significantly, but that we need to wait for the shock to penetrate the opposite side of the companion to obtain precise amounts. It can also be observed that the unbound masses evaluated by both criteria do converge to the same asymptotic value at sufficiently later periods.

\begin{figure}
  \plotone{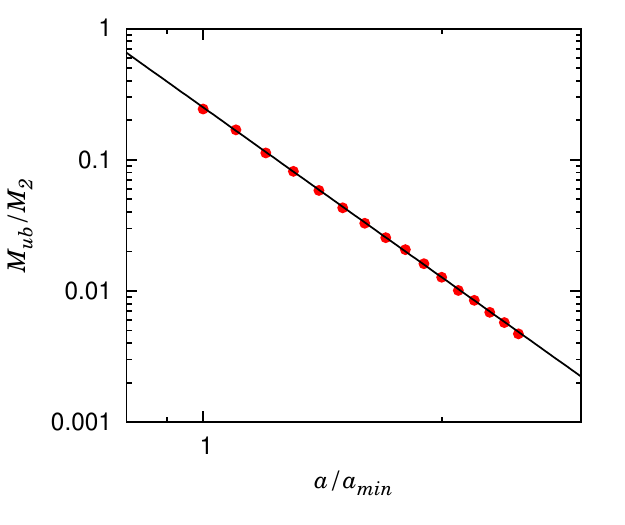}
  \caption{Removed mass as a function of binary separation. Red dots indicate the numerical results. The solid black line is a fit for the results ($\propto a^{-4.3}$).\label{sepdep}}
 \end{figure}

We plot the removed mass as a function of separation in Figure \ref{sepdep}. The red circles indicate the results obtained here. Consistent with the previous studies on type Ia SNe in binaries, the removed mass decreases as the binary separation increases, following the 
power law:
 \begin{eqnarray}
  \frac{M_{\rm ub}}{M_2}=C_{ab}a^{m_{ab}},
 \end{eqnarray}
where $M_{\rm ub}$ is the unbound mass, $M_2$ the companion mass, $C_{ab}$ and $m_{ab}$ fitting parameters.
The solid line in Figure \ref{sepdep} is the best fit for the models with close separations and has $C_{ab}=0.26$ and $m_{ab}=-4.3$. The absolute value of $m_{ab}$ is slightly larger than the values $-3.66$ to $-2.92$ previously
obtained by \cite{pan12} for main sequence and helium star companions, and $-0.416$ to $-0.391$ for red giant companions.

\section{Discussions}

As an attempt to understand the value of the parameter $m_{ab}$, we conducted several experimental calculations varying several physical parameters of SNE. As mentioned already, 
the quantities that change with the binary separation are the solid angle of the companion, the density and pressure of SNE, and the time until the bulk of SNE flows past the 
companion star. The mass and energy contained in the solid angle also vary as a result of these changes. Since kinetic energy dominates in SNE, we ignore the dependence on 
the incident pressure and internal energy in the following.

 We first pay attention to the incident density. As the binary separation increases, it decreases as $a^{-3}$. Since the solid angle of the companion also decreases as $a^{-2}$, the
mass and total energy contained in the solid angle would be reduced even if the density were unchanged, which could have a big impact on the mass loss. In order to see the dependence 
of the removed mass on the incident density alone, though, we artificially amplified the density of the SNE by a factor of 1.2, 1.5, 1.8 and 2.0, without changing other physical 
values including separation. Note that this modification still increases the kinetic energy as well as the momentum of the SNE that collide with the companion. These experiments were 
carried out for models with separations of $(1.0, 1.3, 1.5, 1.8, 2.0) \times a_{\rm min}$.

\begin{figure}
 \plotone{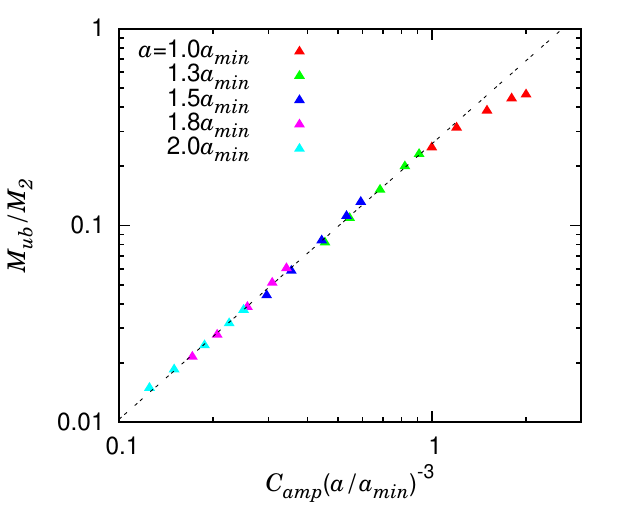}
 \caption{Removed mass vs. incident density of SNE. Plots with the same colour are the results of models with the same separation. The dashed line shows a power law with a power-law index of 
 $1.4$.\label{dendep}}
\end{figure}

 Their results are shown in Figure \ref{dendep}, in which the removed masses are plotted against $(C_{\rm amp}(a/a_{\rm min})^{-3}) \propto\rho_{\rm ej}$, where $C_{\rm amp}$ is the amplification factor of density and 
 the last factor $(a/a_{\rm min})^{-3}$ accounts for the decline in density according to expansion of the SNE. It can be seen that almost all models lie on a single line in the $\log-\log$ plot, which has a power law
 index of $1.4$. This suggests that the removed mass depends solely on $\rho_{\rm ej}$ irrespective of separation as
\begin{equation}
 M_{\rm ub}\propto \rho_{\rm ej}^{1.4}.  
 \label{eq:mub_den_relation}
\end{equation}
This means that the mass and/or total energy injected into the solid angle of the companion star are not very important. These results seem to be at odds with the analytical estimates 
$M_{\rm ub} \propto M_{\rm SN}a^{-2}$ by \cite{wlm75}, in which $M_{\rm SN}$ is the total ejecta mass and corresponds to $C_{\rm amp}$ in our case. Models with the highest $\rho_{\rm ej}$ start to depart 
from the power law. This is because the removed mass becomes comparable to the remaining mass, thus modifying the escape velocity significantly.

Next we artificially multiplied $v^2$ by a factor of $1.2, 1.5, 1.8, 2.0$ as we did with density. The binary separation was fixed to $a = 1.3 \times a_{\rm min}$ this time, implying that the solid angle of 
the companion and hence the density of SNE are unchanged but the incident kinetic energy is still increased, which is expected to have a similar effect to the enhancement of density.
We also tried different combinations of these two effects so that the kinetic energy of the entire SNE would be fixed.

 \begin{table}
  \centering
   \caption{Results for the Experimental Computations\label{phystests}}
   \begin{tabular}{ccccc}
    \hline\hline
    model & $\rho_{\rm ej}$\tablenotemark{a} & $v^2_{\rm ej}$ \tablenotemark{a}& $E_{\rm ej}$\tablenotemark{a} & $M_{\rm ub}(\msun)$\\ \hline
    base  &     1.00    &    1.00    &   1.00   &   0.866 \\
    dd1.2 &     1.20    &    1.00    &   1.20   &   1.107 \\
    dd1.5 &     1.50    &    1.00    &   1.50   &   1.578 \\
    dd1.8 &     1.80    &    1.00    &   1.80   &   2.069 \\
    dd2.0 &     2.00    &    1.00    &   2.00   &   2.404 \\
    vd1.2 &     1.00    &    1.20    &   1.20   &   1.136 \\
    vd1.5 &     1.00    &    1.50    &   1.50   &   1.659 \\
    vd1.8 &     1.00    &    1.80    &   1.80   &   2.212 \\
    vd2.0 &     1.00    &    2.00    &   2.00   &   2.592 \\
    ef1.2 &     1.20    &    0.83    &   1.00   &   0.798 \\
    ef1.5 &     1.50    &    0.67    &   1.00   &   0.770 \\
    ef1.8 &     1.80    &    0.56    &   1.00   &   0.746 \\
    ef2.0 &     2.00    &    0.50    &   1.00   &   0.731 \\
    \hline
   \end{tabular}
   \footnotetext{$\rho_{\rm ej},v^2_{\rm ej}$ and $E_{\rm ej}$ are the values of the SNE normalized by the values of the base model but with $a=1.3\times a_{\rm min}$.}
 \end{table}

The results for these tests are summarized in Table \ref{phystests}. It is evident from the vd-series, in which only the velocity was changed, that the removed mass increases as the velocity 
rises just as expected. Interestingly, the comparison between the vd-series and the dd-series, in which we varied density alone, reveals that if the change in kinetic energy is the same, 
changing velocity has a slightly larger effect on the amount of removed mass than changing density. This is also vindicated by the ef-series, in which both density and velocity were modified but
the kinetic energy was fixed. The latter result may imply that increase in the incident momentum decreases removed mass as long as the incident kinetic energy is identical.

 Now we know how the density and/or velocity of SNE affect the removed mass, our next question will be -- why? We have observed that direct mass stripping was extremely inefficient in our
 models because the SNE is directed away from the companion star once the bow shock is formed. Most of the mass loss occurs instead by ablation owing to the shock heating of stellar matter. 
 This means that the energy, not the momentum, imparted to the companion star is most important. Then the shock strength will be the physical parameter that we should look into, since 
 it determines how much the star will be heated.

 To see how the shock strength depends on the incident density, we consider a Riemann problem as presented in Figure \ref{riemannillu}. This is meant to be a very crude approximation to the 
 collision of SNE against the companion star. The left side of the initial discontinuity represents the SNE, taking typical values of pressure and velocity from our simulation ($p_1=1\times 10^{-2}, 
 v_1=3\times 10^8$). We vary $\rho_1$ to see how this affects the strength of the forward shock produced by collision.  For the right state approximating the envelope of the companion star, 
 we use average values of density and pressure from our companion model: $\rho_2=1\times 10^{-8},p_2=1\times 10^6,v_2=0$. Here CGS units are adopted. We assume that the companion 
 star is at rest. We employ the same assumption as in \cite{che74} to determine the maximum value of $\rho_1$: the ejecta mass is distributed in a uniform shell with a thickness of $\frac{1}{3}$ 
 the shock radius. This gives us the maximum density as a reference:
 \begin{eqnarray}
  \rho_{\rm ej}^{max}=\rho_{\rm ej}(a_{\rm min})=\frac{27}{19}\frac{3M_{\rm ej}}{4\pi a_{\rm min}^3},
 \end{eqnarray}
where $M_{\rm ej}$ is the mass of SNE. 

 \begin{figure}
  \centering
   \begin{picture}(200,60)
    \put(100,0){\line(0,1){60}}
    \put( 50,50){Ejecta}
    \put(125,50){Star}
    \put( 10,35){$\rho_1$:variable}
    \put( 10,20){$p_1=1\times 10^{-2}$erg/cm$^3$}
    \put( 10, 5){$v_1=3\times 10^8$cm/s}
    \put(110,35){$\rho_2=1\times 10^{-8}{\rm g/cm^3}$}
    \put(110,20){$p_2=1\times 10^6$erg/cm$^3$}
    \put(110, 5){$v_2=0$ cm/s}
   \end{picture}
   \caption{Illustration of the Riemann problem.}
   \label{riemannillu}
 \end{figure}

\begin{figure}
   \centering
    \plotone{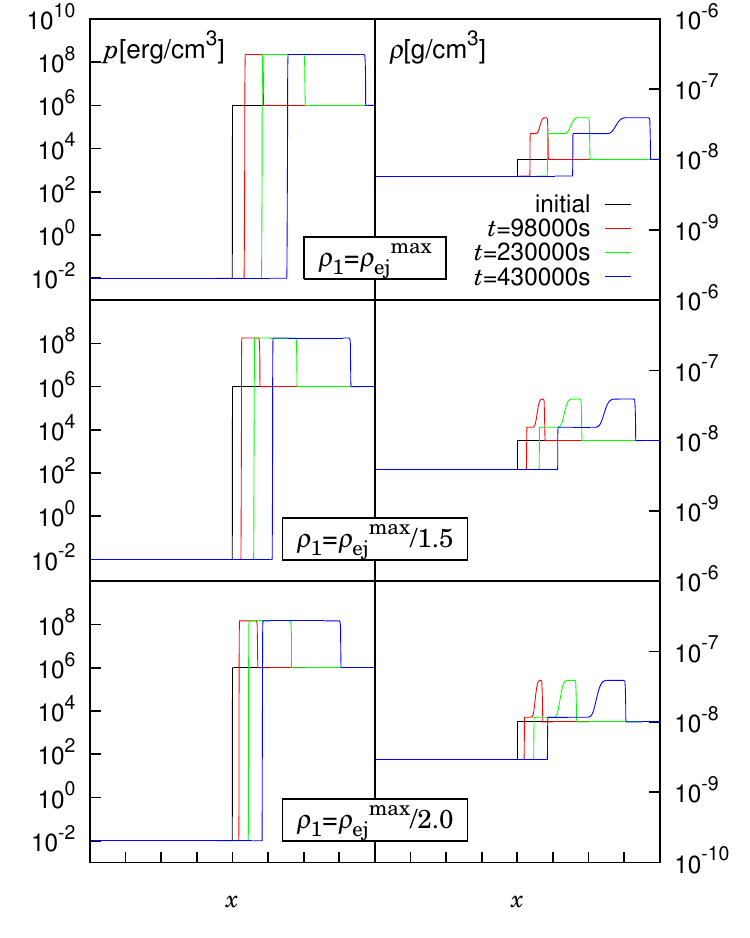}
    \caption{Solutions of the Riemann problem given in Figure \ref{riemannillu}. Left panels show distributions of pressure while right panels display those of density. {\it Top} : results for 
    $\rho_1=\rho_{\rm ej}^{max}$, {\it Middle} : $\rho_1=\rho_{\rm ej}^{max}/1.5$  and {\it Bottom} : $\rho_1=\rho_{\rm ej}^{max}/2.0$. Different lines correspond to different times.\label{riemannevo}}
  \end{figure}

\begin{figure}
   \plotone{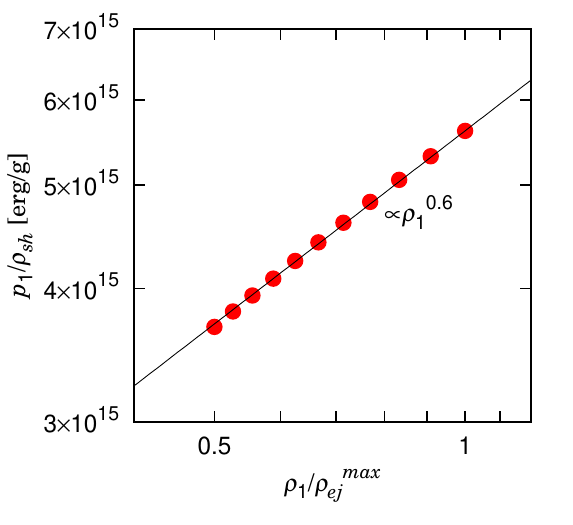}
   \caption{Pressure of the shock-heated region as a function of $\rho_1$.\label{riemanndep}}
  \end{figure}

 The solutions of the Riemann problem are given in Figure \ref{riemannevo}. The left panels show the pressure distributions and the right panels show the density counter part. 
 Each row hold results for different initial conditions, i.e. $\rho_1=\rho_{\rm ej}^{max}$ for the top panels, $\rho_1=\rho_{\rm ej}^{max}/1.5$ for the middle panels, and 
 $\rho_1=\rho_{\rm ej}^{max}/2.0$ for the bottom panels. The different colours of lines correspond to different times. We can see jumps in density and pressure at the positions of the 
 forward shock propagating through the companion star and sweeping up mass. There is also a reverse shock propagating away from the companion into the SNE, which tries to push 
 it back toward the star. A contact surface divides the SNE from the companion and density but not pressure has a jump there. The region between the forward and reverse shocks 
 corresponds to the matter heated by one of these shocks. Figure \ref{riemanndep} shows the pressure divided by the post-shock density $\rho_{sh}$, a measure of the shock strength, as a function 
 of $\rho_1$. We again find a power law, whose exponent is rather sensitive to the value of $\rho_2$.

In our simulations, the forward shock initially has a rather flat front perpendicular to the symmetry axis. It maintains a nearly constant speed until the reverse shock starts to expand toward 
the primary star. This launch of the reverse shock is accompanied by the generation of a rarefaction wave, which in turn propagates towards the forward shock. It eventually catches up and 
sucks the energy that would otherwise be injected to the shock. It is hence natural to expect that the total energy given to the companion is determined at this point. 

\begin{figure}
 \plotone{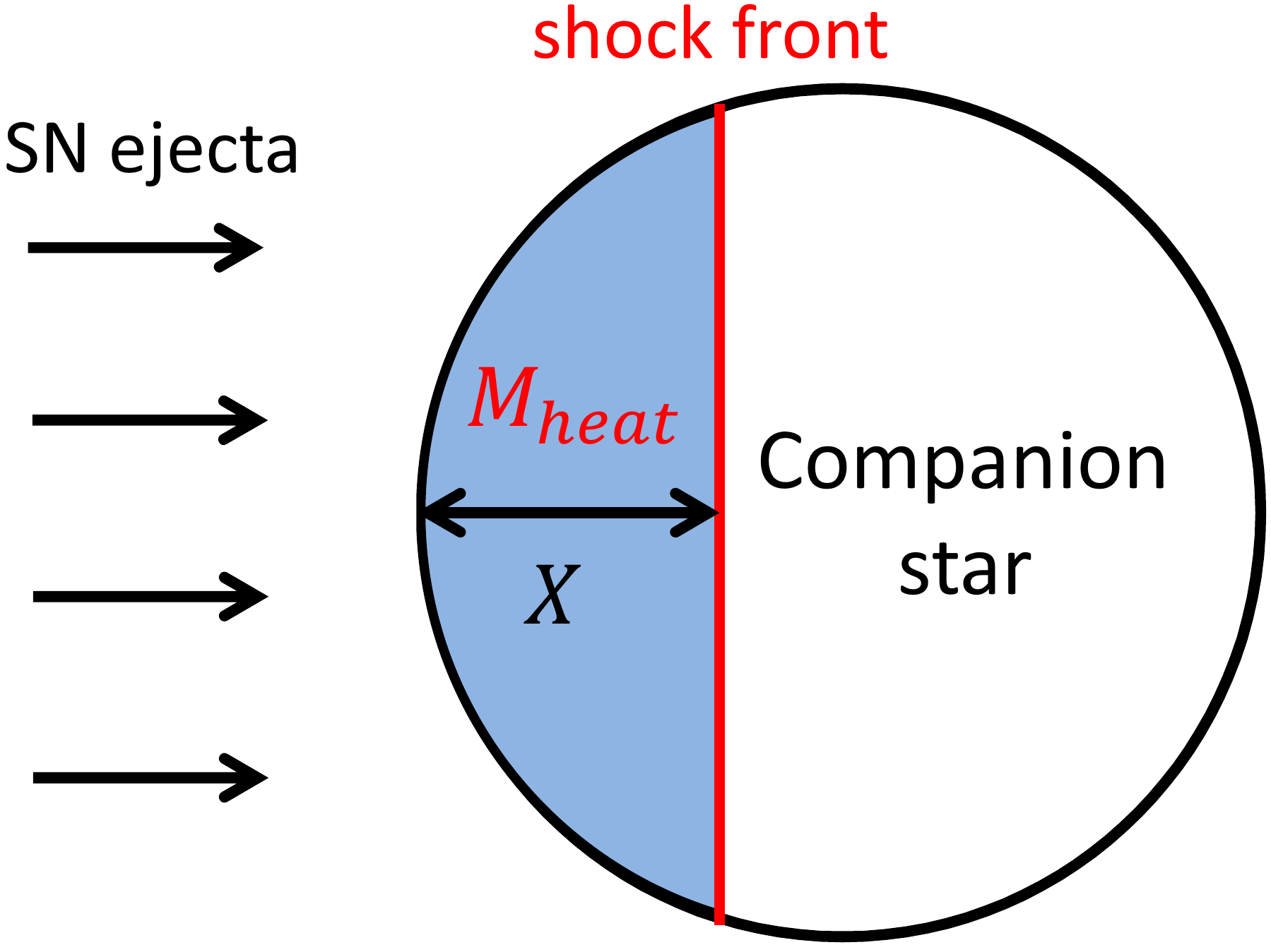}
 \caption{Schematic picture of shock propagation through the companion star. The shaded region is the shock-heated material, whose mass is referred to as $M_{heat}$. The distance from 
 the shock front to the stellar surface is denoted by $X$.\label{mheat}}
\end{figure}

\begin{figure}
 \plotone{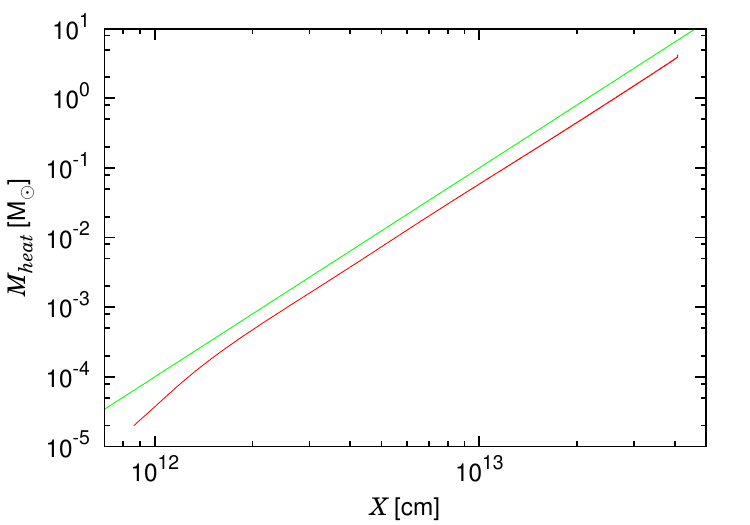}
 \caption{$M_{heat}$ as a function of $X$ (red line). The green line is a power law $X^3$.\label{slicemass}}
\end{figure}

Based on the above observation, we estimate the removed mass as follows. Figure \ref{mheat} displays a schematic picture that gives the idea and notations we use here. Assuming that 
the shock front is a plane perpendicular to the symmetry axis, we approximate the shock-heated portion of the companion star by the shaded region in the picture. $X$ is the distance 
from the stellar surface to the shock front and $M_{heat}$ is the mass contained in this shocked region. We show in Figure \ref{slicemass} $M_{heat}$ as a function of $X$ calculated as
\begin{eqnarray}
 M_{heat}(X)=\int_{r\cos\theta>X}\rho(r)dV.
\end{eqnarray}
 It is probably by coincidence that this function can be fit by a simple power law very well, which allows us to approximate  $M_{heat}$ as
\begin{eqnarray}
 M_{heated}\propto X^3.
\end{eqnarray}

If we assume that the shock propagates at a constant velocity, we have $X=v_{sh}t$, in which $t$ is the time after collision. As argued earlier, the main shock heating stops at  $t=t_{heat}$, 
when the reverse shock starts to expand outwards. We find in our models that $t_{heat}$ roughly corresponds to the time, at which the instantaneous unbound mass reaches the maximum 
(see Figure \ref{timemub}), and that this time does not differ so much among models. Hence assuming that $t_{heat}$ is constant, we obtain $M_{heat}\propto v_{sh}^3$ at $t=t_{heat}$.

The total energy, $E_{heat}$, deposited by the forward shock is estimated as follows, 
\begin{eqnarray}
 E_{heat} &=& \epsilon_{sh}\cdot M_{heat}, \nonumber\\
 &\propto&\rho_{\rm ej}^{0.6}\cdot v_{sh}^3, \nonumber\\
 &\propto&\rho_{\rm ej}^{0.6}\cdot (\rho_{\rm ej}^{0.3})^3, \nonumber\\
 &\propto&\rho_{\rm ej}^{1.5},
\end{eqnarray}
where $\epsilon_{sh}$ is the specific internal energy and is assumed to be proportional to $p_1/\rho_{\rm sh}$;  in the second line we employed the relations $M_{heat}\propto v_{sh}^3$ just 
obtained and $p_1/\rho_{\rm sh} \propto \rho_{\rm ej}^{0.6}$ in Figure \ref{riemanndep}. In the third line, on the other hand, we made use of the approximation $v_{sh}^2 \propto p_1/\rho_{\rm sh}$.
This energy is redistributed in the removed mass via hydrodynamical interactions. It is hence natural to assume that the removed mass is proportional to $E_{heat}$. This hypothesis
finally leads us to the relation $M_{\rm ub} \propto \rho_{\rm ej}^{1.5}$, which is pretty close to the power law with $m_{ab} \sim1.4$ (see Eq.(\ref{eq:mub_den_relation})) obtained in the simulations.

In our simulations, we cut out the central core of the companion star. Although this makes it difficult to accurately estimate the kick velocity imparted by the SNE, it is still possible to make rough estimations on this so called ``rocket effect'' \cite[]{che74}. We calculate the imbalance between the total momentum that flows in from the ejecta side and the momentum that flows out of the other side and find that it is $\sim2\times10^{38} {\rm g\cdot km s^{-1}}$ for the base model with $a=a_{\rm min}$. This should be equal to the momentum deposited to the remnant of the companion star. The kick velocity obtained this way is $v_{kick}\sim2\times10^{38}/(M_2-M_{\rm ub})\sim100$ km s$^{-1}$, which is comparable to the orbital velocity $\sim 50$ km s$^{-1}$ and could destroy the binary system. Note, however, this will be an upper limit, and it decreases as $\propto a^{-2}$ with the binary separation according to the solid angle of the companion.

\section{Summary}

 We have performed hydrodynamical simulations of SNe in binary systems. The removed mass can reach up to approximately $25\%$ of the original mass, and decreases according to a power law of the binary separation. The power is determined by the incident density, which decides the strength of the bow shock.

 The effect of SNE hitting the companion star can become non-negligible at close separations. Such close separations can be created by common envelope evolution, or mass accretion from a tertiary star \cite[]{dvr13}. It is also known from observation that the orbital periods of massive binaries range $\sim1$ day to $\sim3000$ days, with a preference for short periods, meaning that binaries with close separations are popular \cite[]{san12}. When such mass is removed from the companion star, it could alter the evolution from then on. It could change the type of SN if the companion is large enough to cause CCSNe, or could induce kicks.  

Simulations with more realistic binary models will be useful for future detections of surviving companions in SNe such as iPTF 13bvn, SN1993J etc. Observations of SN1993J suggest that the companion star was a B type star \cite[]{mau04,fox14}, which is more compact compared with the companion model in this study. Mass stripping off this compact companion star will surely be less efficient if placed at the same separation as in our simulations. This system may have been fairly close, though, since the companion must have a smaller stellar radius. Since the removed mass strongly depends on incident density and/or velocity rather than the $a/R$ ratio, more mass might have been stripped off this companion by SNE despite its compactness.

 We have not studied the further evolution of the binary as a system, in our simulations. The sudden mass loss should lead to high eccentricities, or could destroy the system. But if the system survives, the expansion of the companion may engulf the remaining neutron star, leading to common envelope evolution \cite[]{liu13-2}. This could create even closer separations, possibly forming HMXBs. After a short time, the secondary star will reach its end of evolution, and cause the second SN. If it is possible for the system to survive this explosion, it may lead to the formation of NS-NS binaries or NS-BH binaries. This study may be the first step to possibly reveal some necessary conditions in forming these compact binaries.

\acknowledgements
 This study was supported by the Grants-in-Aid for the Scientific Research (A) (NoS. 24244036, 24740165), the Grants-in-Aid for the Scientific Research on Innovative Areas, ``New Development in Astrophysics through multi messenger observations of gravitational wave sources'' (No. 24103006).

\clearpage

\end{document}